# Overview of C-ITS Deployment Projects in Europe and USA

Areti Kotsi[1], Evangelos Mitsakis[1], Dimitris Tzanis[1]

[1]Centre for Research & Technology - Hellas (CERTH) - Hellenic Institute of Transport (HIT)
E-mail: akotsi@certh.gr, emit@certh.gr, dtzanis@certh.gr

**Abstract**

Cooperative Intelligent Transportation Systems (C-ITS) are technologies that enable vehicles to communicate with each other and with the road infrastructure. These innovative technologies enable road users and traffic managers to share useful information, assisting the coordination of their actions. During the last years various initiatives providing policy rules for C-ITS deployment and a large number of projects demonstrating C-ITS implementation have taken place in Europe and USA. However, the identification of the status of C-ITS deployment remains ambiguous at binational level. The purpose of this paper is to provide an overview of the European and US milestones, that have been reached so far in the field of C-ITS, by identifying and reporting the policy framework, as well as the projects concerning C-ITS deployment in Europe and USA.

**Περίληψη**

Τα Συνεργατικά Ευφυή Συστήματα Μεταφορών (Σ-ΕΣΜ) αποτελούν τεχνολογίες οι οποίες επιτρέπουν την επικοινωνία μεταξύ οχημάτων και μεταξύ οχημάτων και υποδομής. Η συγκεκριμένη καινοτόμα δέσμη τεχνολογιών δίνει τη δυνατότητα στους χρήστες του οδικού δικτύου και στους διαχειριστές της κυκλοφορίας να ανταλλάσσουν χρήσιμες πληροφορίες με αποτέλεσμα τον καλύτερο συντονισμό των δράσεών τους. Τα τελευταία χρόνια εχουν θεσπιστεί διάφορα κανονιστικά πλαίσια για την ανάπτυξη των Σ-ΕΣΜ και έχουν πραγματοποιηθεί πολλά έργα με στόχο την επίδειξη εφαρμογών Σ-ΕΣΜ. Ωστόσο η ακριβής κατάσταση γύρω από την ανάπτυξη των Σ-ΕΣΜ παραμένει ακόμη ασαφής σε διεθνικό επίπεδο. Ο σκοπός της παρούσας εργασίας είναι να πραγματοποιήσει μια επισκόπηση γύρω από τα ορόσημα που έχουν σημειωθεί στην ανάπτυξη των Σ-ΕΣΜ στην Ευρώπη και στις Ηνωμένες Πολιτείες, παρουσιάζοντας πολύτιμες πληροφορίες για το κανονιστικό πλαίσιο και τα έργα που αφορούν στην ανάπτυξη των Σ-ΕΣΜ στην Ευρώπη και στις Ηνωμένες Πολιτείες.

*Λέξεις-κλειδιά: Συνεργατικά Ευφυή Συστήματα Μεταφορών, ανάπτυξη ευρείας κλίμακας, καινοτομία, έργα.*

## *1. Introduction*

The technological progress of the last decade led to the expansion of transportation research especially in the field of Cooperative Intelligent Transportation Systems (C-ITS). This paper aims to provide the reader with information about the status of the deployment of C-ITS in Europe and USA. The reader is first introduced to European and US policies concerning C-ITS, while a detailed presentation of C-ITS deployment projects in Europe and USA constitutes the main part of this paper.

The European C-ITS policy initiatives referring to the timeframe of 2010-2017 are: the Directive 2010/40/EU (European Commission, 2010), the 2011 White Paper "Roadmap to a Single European Transport Area - Towards a competitive and resource efficient transport system" (European Commission, 2011), the "Roadmap between automotive industry and infrastructure organisations on initial deployment of Cooperative ITS in Europe" (Amsterdam Group, 2013), the list of "Day 1 applications" (European Commission, 2016a), the "AG White Papers" (Amsterdam Group), the document "C-ITS Platform - Final Report - January 2016 (Phase 1)" (C-ITS Platform, 2016), the Roadmap "A Master Plan for the deployment of Interoperable Cooperative Intelligent Transport Systems in the EU" (C-ITS Master Plan) (European Commission 2016b), the "Declaration of Amsterdam - Cooperation in the field of connected and automated driving on connected and automated driving" (European Commission 2016c), and the "C-ITS Certificate Policy for deployment and operation of European C-ITS" (European Commission, 2017a).

US C-ITS policy initiatives covering the time period between 1996-2017 comprise of: the "National ITS Architecture" (US Department of Transportation, 2019a), the "Turbo Architecture" (US Department of Transportation, 2019b), the standards for CV deployment by the US Department of Transportation Intelligent Transportation Systems Joint Program Office (USDOT ITS JPO), the agreement "EU - U.S. Joint Declaration of Intent on Research Cooperation in Cooperative Systems" (Stančič Z., Appel. P. H., 2009), the guideline "2015 FHWA Vehicle to Infrastructure Deployment Guidance and Products" (US Department of Transportation, 2014), the system architecture framework "Connected Vehicle Reference Implementation Architecture" (CVRIA) (US Department of Transportation, 2017a), the "Architecture Reference for Cooperative and Intelligent Transportation" (ARC-IT) (US Department of Transportation, 2017b), and two tools, "Regional Architecture Development for Intelligent Transportation" (RAD-IT) (US Department of Transportation, 2017c) and "Systems Engineering Tool for Intelligent Transportation" (SET-IT) (US Department of Transportation, 2017d).

## *2. Large-scale deployment projects in Europe and USA*

### *2.1 European activities*

The European Commission (EC) launched under the Sixth Framework Programme (FP6) (2002-2006) the projects SAFESPOT (European Commission, 2007), COOPERS (European Commission, 2008) and CVIS (European Commission 2016d), all focusing on the provisions of real-time safety related traffic/ infrastructure information for drivers. The following funding programme, 7th Framework Programme for Research and Technological Development (FP7) (2007-2013), provided the budget for a series of research and innovation initiatives, including the projects: SAFERIDER (European Commission, 2017b), euroFOT (European Commission, 2017c), SISCOGA (FOT-NET DATA), PRESERVE (European Commission, 2017d), COMeSafety2 (European Commission, 2017e), FOTsis (European Commission, 2017f), ITSSv6 (European Commission, 2017g), MOBiNET (European Commission, 2017h), P4ITS (European Commission, 2017i), VRUITS (European Commission, 2014a), COMPANION (European Commission, 2014b), HeERO2 (European Commission, 2017j), TeleFOT (European Commission, 2017k), DRIVE C2X (European Commission, 2014c), eCoMove (European Commission, 2017l), interactIVe (European Commission, 2017m), and OVERSEE (European Commission, 2017n). The objectives of the projects covered various aspects concerning cooperative systems, such as the provision of Advanced driver-assistance systems (ADAS) and In-vehicle infotainment (IVI) systems, the assessment of C-ITS through Field Operational Tests (FOTs) in various places in Europe, the provision of security and privacy subsystems for Vehicle-to-Everything (V2X) communication systems, the development of open platforms for Europe-wide mobility services, as well as the development of architectures for C-ITS services.

During the same period (2007-2013), the Competitiveness and Innovation Framework Programme (CIP) funded five projects. The FREILOT project aimed to increase energy efficiency in road goods transport through C-ITS services (European Commission, 2017o), while the COSMO project demonstrated the energy efficiency-related benefits of integrating advanced cooperative traffic management systems (European Commission, 2017p). The HeERO project prepared the deployment of the necessary infrastructure in Europe for the "Pan-European in-vehicle emergency call service, eCall" (European Commission, 2017q).

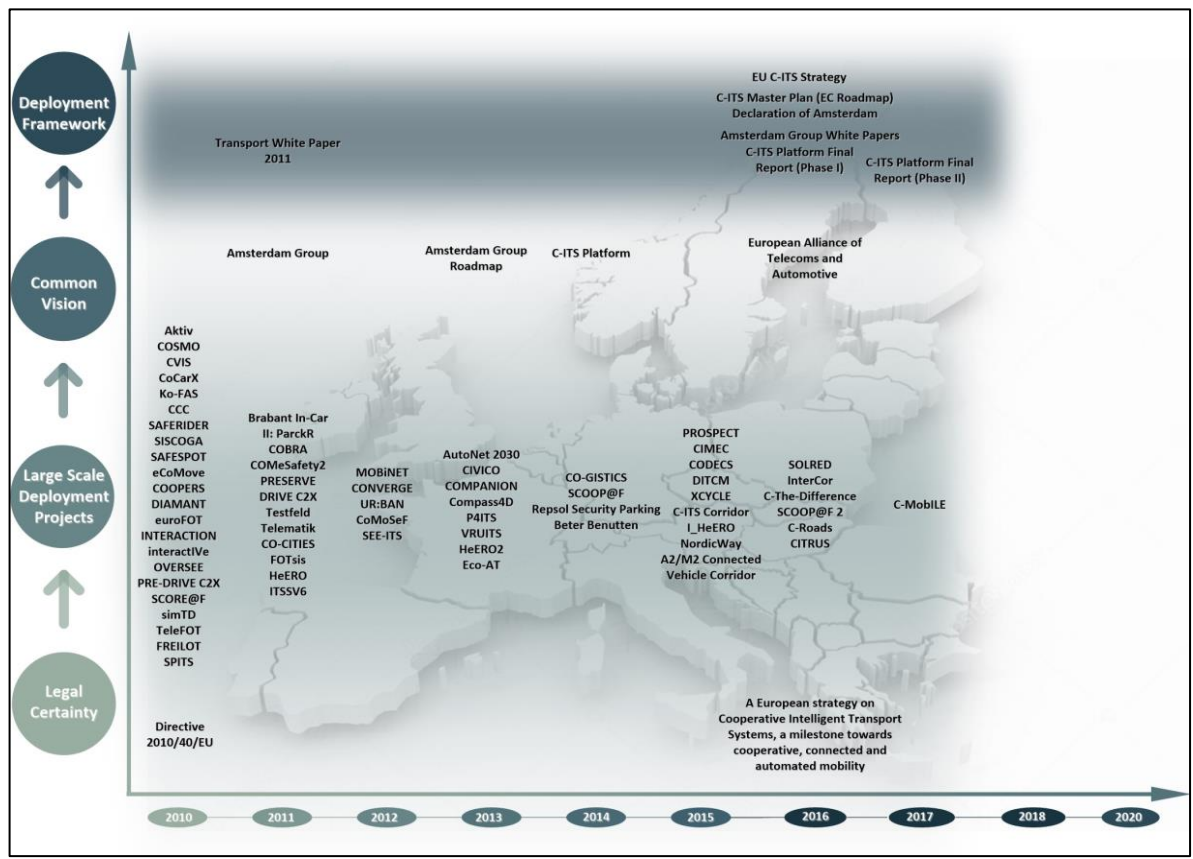

*Figure 1: C-ITS European activities overview*

Through the Compass4D, three cooperative services (Road Hazard Warning, Red Light Violation Warning and Energy Efficient Intersection service) were implemented in seven European cities (Bordeaux, Copenhagen, Eindhoven-Helmond, Newcastle, Thessaloniki, Verona and Vigo) (European Commission, 2017r). The CO-GISTICS project aimed to implement C-ITS services, which would increase energy efficiency and equivalent $CO_2$ emissions, in seven European cities/ logistics hubs (Bordeaux, Frankfurt, Thessaloniki, Trieste, Arad, Bilbao and Vigo) (European Commission, 2017r).

The EU funding instrument Connecting Europe Facility (CEF) (2014-2020) supported a significant number of C-ITS activities, including the projects SCOOP@F (European Commission, 2015), Repsol Security Parking (European Commission, 2018a), I_HeERO (European Commission, 2018b), NordicWay (European Commission, 2018c), CITRUS (European Commission, 2009), SolC-ITS (European Commission, 2018d), InterCor (European Commission, 2018e), and C-Roads (C-ROADS). The focus of the projects is mainly on the establishment of a harmonized strategic rollout and common specifications on C-ITS implementation among EU members, aiming in the promotion of wide-scale C-ITS deployment in Europe. Under the EU Research and Innovation programme Horizon 2020 (2014-2020), the following projects were launched: PROSPECT (European Commission, 2017t), CIMEC (European Commission, 2017u), CODECS (European Commission, 2017v), XCYCLE (European Commission, 2017w) and C-MobILE (European Commission, 2018f). Fields covered under the framework of these projects comprise of the support of strategic and technical policy solutions and processes for a consolidated C-ITS rollout, as well as the deployment of C-ITS services in urban environments, with the scope of establishing a fully safe and efficient road transport for Vulnerable Road Users (VRUs).

A list of projects funded under the funding programmes ICT Policy Support Programme (ICT PSP), ERA-NET, South East Europe Transnational Cooperation Programme, Celtic-Plus, European Regional Development Fund (ERDF), and EC DG MOVE are respectively: the COBRA project (Vermaat P., Hopkin J., van Wees K. A. P. C., Faber F., Deix S., Nitsche P., and Michael. K., 2012), the SEE-ITS project (Mitsakis E., Iordanopoulos P., Aifadopoulou G., Tyrinopoulos Y., and Chatziathanasiou M., 2012), the CoMoSeF project (Celtic-Plus Smart Connected World), and the project C-The Difference (Blervaque V., 2016). At national level, Germany introduced the projects CoCarX (Cooperative Cars eXtended – CoCarX, 2011), simTD (FOT_NET DATA), Ko-FAS (European Commission, 2018h), and UR:BAN (UR:BAN Urbaner Raum: Benutzergerechte Assistenzsysteme und Netzmanagement), focusing on Car-to-Car (C2C) and Car-to-Infrastructure (C2I) communication for future C-ITS applications. The Netherlands launched the projects SPITS (FOT-NET DATA), Brabant In-Car II: ParckR (FOT-NET DATA), Beter Benutten (Optimising Use) programme (Ministry of Infrastructure and the Environment, 2017) and DITCM (DITCM Innovations, 2017). Austria through the Testfeld Telematik project, aimed for "developing, operating and demonstrating C-ITS services and systems within the framework of a test field in the greater Vienna area" (Eco-AT The Austrian contribution to the Cooperative ITS Corridor), while the ECo-AT project intended to "create harmonized and standardized C-ITS applications jointly with partners in Germany and the Netherlands" (Eco-AT The Austrian contribution to the Cooperative ITS Corridor). The French FOT, titled SCORE@F, implemented ITS stations, which used a standardized communication architecture (ITS Station Reference Architecture) (FOT-NET DATA), while UK launched the A2/ M2 (London to Dover) Connected Vehicle Corridor project, in order to create a living laboratory for C-ITS technologies (Department of Transport, 2016). At international level, German, Dutch and Austrian road operators set the basis for "a European-wide C-ITS implementation" under the Cooperative ITS Corridor project. The corridor Rotterdam-Frankfurt/M.-Vienna includes the implementation of two C-ITS services, Road Works Warning and Vehicle Data for improved traffic management (Cooperative ITS Corridor Joint deployment).

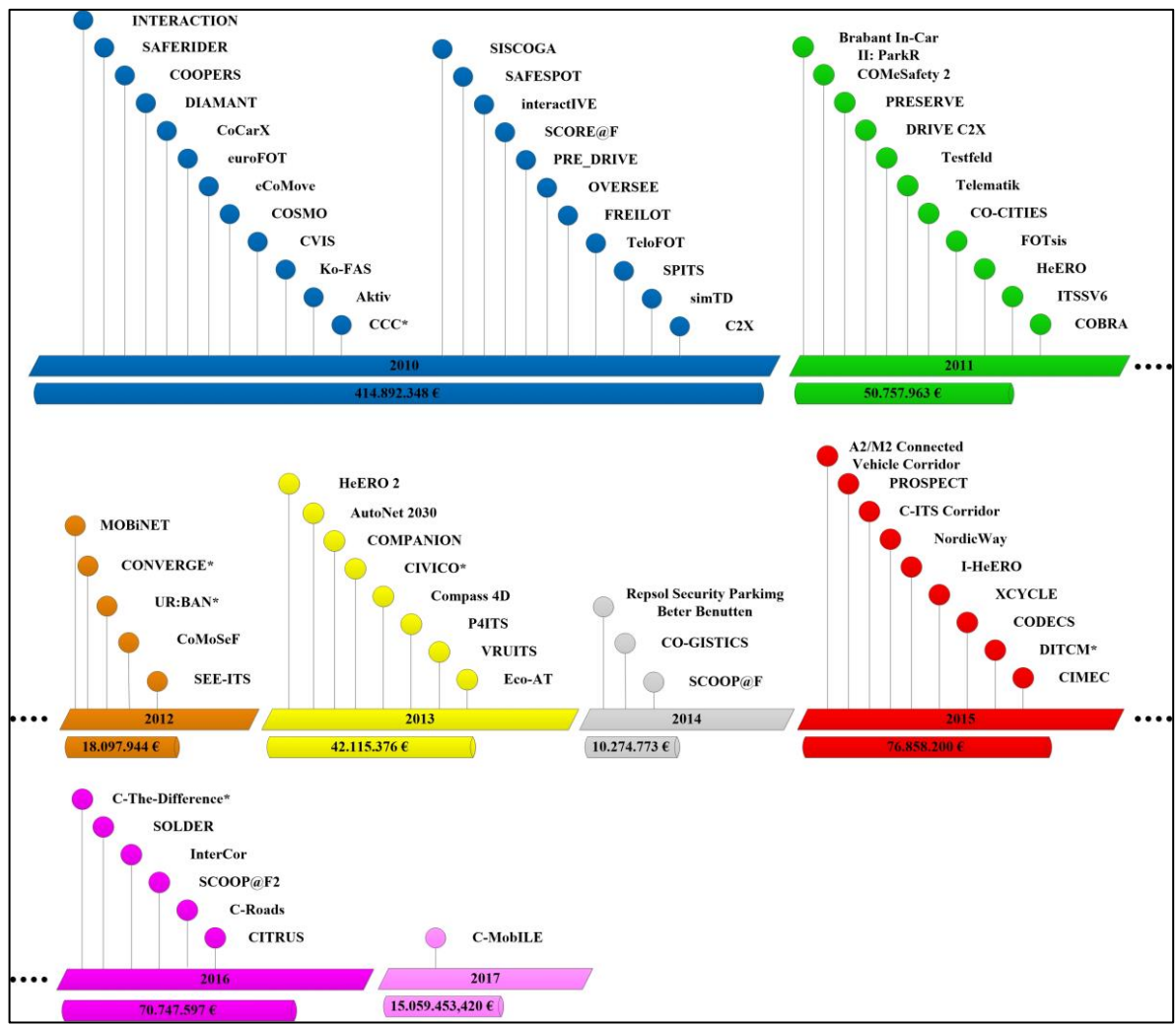

*Figure 2: Large-scale deployment projects in Europe*

## 2.2 USA activities

In 2008, the Arizona Transportation Research Center, the Arizona State University, the Maricopa County DOT and the Michigan DOT funded the pilot project "Arizona Emergency Vehicle Infrastructure Integration (E-VII)". The project consisted of two phases: Phase 1, including the analysis and deployment of prototype applications, and Phase 2, including the demonstration of applications, equipment interfaces and driver interaction with the on-board systems in a "parking lot" site at the Maricopa County. The scope was the development and testing of advanced technologies for emergency vehicles, providing better response to traffic incidents (Saleem F., and Nodes. S., 2008). In 2009, the Michigan DOT, the University of Michigan Transportation Research Institute (UMTRI) and the Institute for Information Industry launched the "Multipath Signal Phase and Timing (SpaT) Broadcast project". The aim of the project was the provision of speed advice to drivers, in order to pass safely through the green phase of the following signalized intersection. Moreover, SPaT provided drivers with a down counter, demonstrating via the vehicle's interface the remaining seconds of the green phase (Robinson R., and Dion F.).

In 2011 UMTRI in cooperation with USDOT, introduced the "Connected Vehicle Safety Pilot Model Deployment (SPMD)" project. The project included the collection of real-time data, in order to evaluate the effectiveness of Connected Vehicles (CV) safety technologies. Over 2800 volunteer vehicles from Ann Arbor were equipped with Vehicle-to-Vehicle (V2V) and Infrastructure-to-Vehicle (I2V) communication devices, in order to exchange Basic Safety Messages (BSMs) about speed, location and direction in a 73 lane-miles equipped roadway (US Department of Transportation, 2018a). In the same year, USDOT, Michigan DOT and the Federal Highway Administration (FHWA) sponsored the "Integrated Mobile Observations 2.0 (IMO)" project. The project implemented a system acquiring weather-road data from the I-94 corridor users (fleet of 60 vehicles) and providing it to weather analysts. The implemented system, an Android-based customized smartphone device (DataProbe), collected all the vehicle data, then transferring it to an UMTRI server for validation, storage and analysis (Belzowski B. M., and Cook. S. J.).

FHWA and Auburn University sponsored the "Heavy Truck Cooperative Adaptive Cruise Control" project, addressing the implementation of the Driver Assistive Truck Platooning (DATP), a form of Cooperative Adaptive Cruise Control (CACC) for heavy trucks. The equipment included radars, dedicated short-range communications (DSRC) based on V2V communications and satellite positioning technologies (Dr. Bevly D., Dr. Murray C., Dr. Lim A., Dr. Turochy R., Dr. Sesek R., Smith S., Apperson G., Woodruff J., Gao S., Gordon M., Smith N., Watts A., Dr. Batterson J., Bishop R., Murray D., Torrey F., Korn A., Dr. Switkes J., and Boyd S., 2015). In 2015 the University of Washington sponsored and coordinated the "Enhancing Safe Traffic Operations using Connected Vehicles Data" project. The project developed a cost-effective Communication Note (CN) device and a mobile application (Android based), in order to advise drivers about hazardous scenarios and to warn VRUs (Li Z., 2016). Expanding the SPMD project, UMTRI and its partners launched the "Ann Arbor Connected Vehicle Test Environment" project, aiming to broaden the existing infrastructure to the entire 27-square miles of the City of Ann Arbor, and to include additional equipped vehicles (5000 until 2018) (UMTRI).

In an effort of expanding CV technology, USDOT launched the "Connected Vehicle Pilot Deployment Program". The main objective of the program was the innovative and cost-effective combination of CV technologies and mobile applications, targeting in traveler mobility and safety increase, and environmental impacts reduction. The first phase of the program, completed in September 2016, included "a concept of operations, system requirements, safety plan, benefits evaluation plan, security management plan and development plan". The second phase, lasting 20 months, embodies "the detailed design, field equipment development and procurement, software development, integration and the installation of the in-vehicle devices and the roadside infrastructure". During the last phase, applications will be active and provided to drivers in New York City (NYC), Tampa (Florida) and Wyoming (US Department of Transportation, 2018b).

The NYC pilot includes three corridors: Manhattan Grid, Manhattan FDR Drive and Brooklyn Flatbush Avenue. The implemented applications focus on safety, promoting NYC's Vision Zero program. The implemented safety applications are based on V2V, I2V and Infrastructure-to-Pedestrian (IVP) communications, covering 300 Road Side Units (RSUs) in the 3 corridors and 8000 CV (US Department of Transportation, 2018c). Tampa pilot focuses on the deployment of V2V and I2V applications, in order to reduce congestion and collisions. Additionally, the pilot aims to "enhance pedestrian safety, speed bus operations and reduce conflicts between street cars, pedestrians and passenger cars at locations with high volumes of mixed traffic" using CV technology. Participants of the pilot deployment comprise of 1600 cars, 10 buses, 10 trolleys, 500 pedestrians, and 40 RSUs (US Department of Transportation, 2018d). The pilot site of Wyoming is a major freight corridor, assisting in the movement of goods across U.S.A. The pilot targets in safety improvement and reduction of "incident-related delays" via the following V2V and I2V applications: Forward Collision Warning (FCW) and Distress Notification. 400 vehicles (150 heavy trucks, 100 fleet vehicles, snowplows and highway patrol vehicles), equipped with On-Board Units (OBUs), as well as 75 RSUs will support operations (US Department of Transportation, 2018e).

In December 2015, USDOT introduced a project called "Smart City Challenge". The purpose of the project was the development of ideas, from mid-sized American cities, in order to establish a new and smart transportation system (US Department of Transportation, 2018f). City of Columbus won the challenge, trying now to revolutionize its transportation system through the "Columbus Smart City demonstration" project. The project includes CV deployment, upgrade of infrastructure and the development of an integrated data platform. Columbus DOT focuses on improving safety, mobility and environmental sustainability via the following CV applications: Transit Signal Priority, Freight Signal Priority, Eco-Approach and Departure at Signalized Intersections, Forward Collision Warning, Emergency Vehicle Preemption, Red Light Warning Violation, Speed Warnings at School Zones, Vehicle Turning Right in Front of Transit Vehicle, and Emergency Electronic Brake Light Warning. 400 city vehicles, 50 trucks, 100 school buses, 350 buses, and 2100 private light vehicles, equipped with DSRC units, as well as 175 intersections, equipped with RSUs, comprise the components of the implementation (US Department of Transportation, 2018g).

In 2016 the Federal Tiger Grant supported the TPIMS project, "Truck Park Information Management Systems". The goal of this project was the provision of real-time information to truck drivers, in order to assist in right and economical parking decisions (Mid America Association of State Transportation Officials, 2016). Within the Ohio Smart Mobility Corridor project, the US Route 33 was equipped with fiber optic cables, enabling researchers and traffic monitors to have a real-time connection with road wireless sensors (Smart Mobility Corridor).

In 2017 the "5.9 GHz Dedicated short-range communication Vehicle-based Road and Weather Condition Application" project was launched, aiming to obtain road-weather information (US Department of Transportation, 2018h).

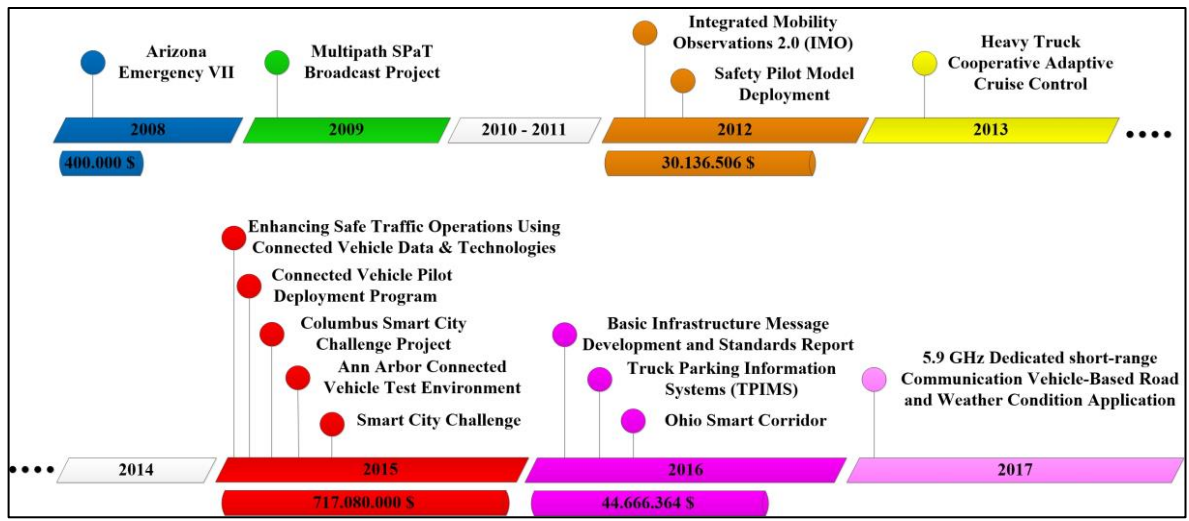

*Figure 3: Large-scale deployment projects in USA*

Regarding Figure. 2 and Figure. 3, the depicted budgets refer only to projects with available total cost information. Projects with unavailable data are not included.

## *3. Conclusions*

The identification of the status of C-ITS deployment projects in Europe and USA demonstrates a large number of activities, devoted to the design and development of C-ITS. Common objective is the integration and wide-scale implementation of C-ITS technologies for the provision and enhancement of road safety, environmental sustainability, as well as efficient mobility of goods and people.

Europe owns a sufficient C-ITS deployment strategy, promoting joint C-ITS deployment activities, which aim to develop and share technical specifications, and to verify interoperability through cross-site testing. Europe's vision on the interoperable deployment of C-ITS is supported by a significant number of C-ITS projects, funded under several programmes, with the scope of creating innovative C-ITS services for all European citizens. USA has proceeded in the development of regulatory and policy rulings, required to foster the growth of CV, in order to tackle big challenges in transportation. The USDOT focuses on coordinating the collaboration amongst various stakeholders to achieve CV deployment and widespread implementation in the real world. Through a series of projects, USDOT supports its first priority, which is to improve roadway safety conditions for American drivers, as well as to exhibit profound effects on VRUs' safety.

Conducting a comparison between the European and US C-ITS deployment activities, it is observed that Europe, although having an adequate deployment background, is still in initial steps for tackling the technical challenges of C-ITS interoperability. On the other hand, USA has managed to establish clear guidelines, and to define C-ITS deployment under a national architecture framework. The future of C-ITS technologies is already coming into focus for both Europe and USA, leading the way in the development of autonomous mobility.

## *4. References-Bibliography*

https://www.its.dot.gov/presentations/Road_Weather2014/6A%20Garrett_CTS_PFS_DSRC_RdWx_20140813.pdf.